\documentclass[twocolumn,showpacs,preprintnumbers]{revtex4} 
\usepackage{amssymb}
\usepackage[dvips]{graphicx}

\begin{document}

\title{What does intrinsic tunnelling spectroscopy really examine?}

\author{V.N. Zavaritsky$^{1,2}$}
\address
{$^{1}$Department of Physics, Loughborough University, Loughborough, United Kingdom, 
$^{2}$Kapitza Institute for Physical Problems and General Physics Institute,  Moscow, Russia\\
}

\begin{abstract}
The out-of-plane current-voltage (I-V) characteristics of Bi2212  are studied in experimental environments of different heat transfer efficiency,  allowing practical separation of intrinsic and extrinsic phenomena. {\it Intrinsic} (heating-free) response is Ohmic in the normal state of Bi2212, while its resistance, R=V/I, is found to be a good practical measure of the mean temperature of the sample in the overheated case. A self-heating model proposed for the latter case provides a qualitative and quantitative description of key findings of  intrinsic tunnelling spectroscopy including (pseudo)gaps, quasiparticle and normal state resistances. The model also naturally explains the  `superconducting' gap closure well below $T_c$ of the material as well as its survival at a magnetic field significantly exceeding $H_{c2}$. The generic shape of the individual branches of the brush-like part of I-V established under conditions of negligible overheating suggests a phase-slip origin of the intrinsic Josephson effect (IJE).
\pacs{74.45.+c, %Proximity effects; Andreev effect; SN and SNS junctions  
 74.50.+r, %Tunneling phenomena; point contacts, weak links, Josephson effects in superconductors  
74.72.-h, %Cuprate superconductors (high-Tc and insulating parent compounds)  
%74.72.Hs, 
74.25.Fy, %Transport properties of superconductors including electric and thermal conductivity, thermoelectric effects, etc  74.25.Fy, %74.25.Op, 74.20.Mn
}
\end{abstract}

\maketitle
\section{Introduction}
 Despite significant efforts to understand the mechanism of high temperature superconductivity (HTSC), no consensus has  been achieved either in theory \cite{theory} or experimentally. In particular, conflicting conclusions on the value and temperature dependences of the superconducting gap, $\Delta_s$, and the normal state pseugo-gap, $\Delta_p$, in the electronic density of states (DOS) have been drawn from conventional spectroscopy \cite{spectroscopy,Miyakawa,Fedorov,Ponomarev,STM-HTS,literature} and from `intrinsic' tunnelling ({\it cf,} \cite{yurgens2201} and references therein). Intrinsic tunnelling (ITs) is believed to occur in highly anisotropic superconductors such as Bi2212 and Bi2201, manifesting itself in the peculiar shape of an out-of-plane I-V (IVC). The low-bias brush-like part of such I-V is highly hysteresial and becomes single-valued at higher voltages, where gap-like peculiarities are located. In resemblance of `conventional' Josephson junctions' I-V, individual branches of that brush were found to be highly nonlinear. 

This similarity has led to a belief that layered HTSC such as Bi2212 represent natural stacks of atomic scale intrinsic superconductor-insulator-superconductor (SIS) - Josephson junctions. Correspondingly, it is generally believed that the c-axis (out-of-plane) charge transport in layered HTSC occurs via a sequential tunnelling of electrons or Cooper pairs between Cu-O planes. Some important conclusions have been drawn from results obtained with this method, where Josephson coupling between Cu-O planes was assumed to be the {\it only} cause of the peculiar shape of I-V. This assumption, however, is not beyond dispute from the experimental viewpoint \cite{comment-1st,heating-1st,comment2201}. In particular, unlike in conventional tunnelling, the characteristic features of ITs are seen at high heat loads, $P$, (estimated as VI/A, where A is the area of the sample) in excess of {\it kilowatts} per cm$^2$ at T$\sim$4.5K. This value significantly surpasses the critical load for liquid $^4$He, $\sim$1W/cm$^2$, thus signalling possible heating, especially since the poor thermal conductivity of all HTSC makes them particularly prone to local overheating. Nevertheless, heating issues in IT studies of HTSC were mostly misinterpreted or ignored until recently, when direct experimental evidence of the heating origin of I-V non-linearities in Bi2212 mesas was obtained \cite{heating-1st}.

The majority of IT investigations use bridge-like structures (mesas) fabricated from HTSC crystal or film. Unlike tunnel junctions, these mesas, like any superconducting weak links, are characterised by a high current density, which generates a nonequilibrium state in the structure \cite{licharev}. The stationary and non-stationary properties of the nonequilibrium state in conventional superconductors have been a subject of systematic investigation. In particular, phase-slip centers (PSC) of different dimensionality have been studied in great detail \cite{phase-slip,Bezryadin,ustinov} as have a rich variety of nonlinear phenomena caused by Joule self-heating. The latter often mimics Josephson-like behaviour \cite{gurevich} so that care must be taken to distinguish intrinsic properties from extrinsic. 

Below I report on an experimental study of c-axis I-V of Bi2212 single crystals and mesa-structures of the lateral in-plane size, W=10-300$\mu m$, performed both under conditions of negligible self-heating and in the opposite scenario,  where I-V characteristics are dominated by Joule self-heating. A self-heating model of IT is proposed on these experimental grounds and is found to provide a quantitative description of IT phenomenon. Examples drawn from the most accurate investigations will show that the majority of results obtained with the IT technique could be reproduced using the temperature dependence of sample resistance in the normal state, $R_N(T)$, and with the assumption that the heating is  the {\it only} cause of non-linearities observed in current-voltage characteristics. I will also briefly discuss heating-free data and possible causes of multiple branching. The technique employed for study of the quasi-stationary I-V, together with some other experimental details, can be found elsewhere \cite{my_i-v}.

\section{results and discussion}
Our discussion begins with the experimental out-of-plane I-V measured under conditions of both negligible self-heating and in the opposite scenario, where Joule self-heating dominates. For the latter case, the self-heating model is proposed. A few examples drawn from reliable publications will then demonstrate that this model self-consistently  explains the most significant findings obtained with IT spectroscopy, including the pseudogap, the superconducting gap and their coexistence, as well as the contrasting effect of magnetic field on $\Delta_s$ and $\Delta_{pg}$. The following sections consider low-bias IT results and some smoking guns for the model; and the last section is devoted to the `heating-free' I-V and the possible cause of the brush-like type of out-of-plane I-V in Bi2212.

\subsection{Experimental basis of the model}
At first, we considered how the heating-free I-V would look. In our studies, we followed several simple guidelines. Any I-V is inevitably accompanied by Joule dissipation which can escape only to the bath. Mean overheating therefore depends on the Joule heat and the efficiency of the heat transfer into the bath, which can be modified significantly. However good the thermal contact, special attention must be paid to the initial (low-bias) part of the characteristics, where overheating is smaller. 

Indeed, the measurements performed at sufficiently small heat loads in the normal state of the Bi2212 crystal were found to be virtually unaffected by self-heating. The linearity of the out-of-plane I-V observed in this case over 4-9 orders of heat load magnitude and reasonable concordance between the threshold values $P_c$, at which the I-V linearity breaks down and where the Joule heating of the crystal becomes measurable, suggest the intrinsic (heating-free) response to be Ohmic in the normal state of Bi2212. 

This approach is not directly applicable below $T_c$ since the superconducting critical current, $I_c$, imposes a lower limit on the measuring current. A non-vanishing $I_c$ often causes transition of the superconductor into a self-sustaining regime in which the Joule heat release becomes sufficient to raise the temperature above $T_c$ \cite{gurevich}. For this reason, below $T_c$ the `heating-free' I-V can only be addressed reliably  under conditions of sufficient $I_c$ suppression. Both temperature and magnetic field make it possible to suppress $I_c$, thereby providing a practical tool to distinguish intrinsic phenomena from self-heating.

In agreement with the earlier findings \cite{heating-1st,gurevich}, it was found that at sufficiently low temperatures self-heating, which may occur at currents much lower than $I_c$, destroys the superconducting state in Bi2212. Moreover, my experiments demonstrate  that I-V similarity with Josephson or tunnel characteristics does not necessarily imply them to be of the same origin. For example, Joule overheating of the crystal is largely responsible for the Josephson-like shape of I-V in Fig.1A. 

\begin{figure}
\begin{center}
\includegraphics[angle=-0,width=0.47\textwidth]{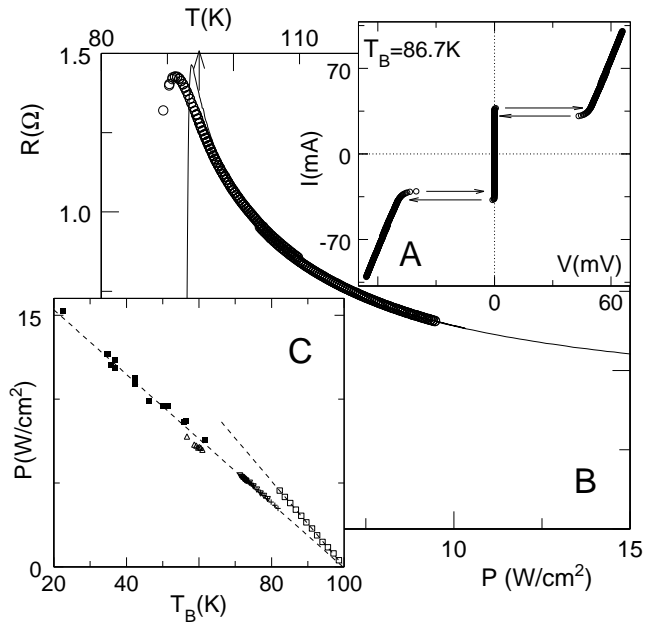}
\vskip -0.5mm
\caption{Inset A shows the typical I-V in the presence of heating. The symbols in {\bf B} present the same I-V,  re-plotted as V/I vs W, together with ac-R(T) of the same sample measured at low bias (solid line, the top x-scale). {\bf C} Heat loads required to heat the crystal from $T_B$ to 100K.}
\end{center}
\end{figure}

In terms of heating issues,  it is appropriate to consider R=V/I as a function of the heat load, $P=VI/A$, rather than I-V only (see \cite{physicaC} for details). The shape of the I-V curves accordingly re-plotted recalls that of the experimental $R(T)$ of the same sample. Moreover, these different dependencies could be superimposed by a simple re-scaling, as shown in Fig.1B, which presents the same I-V properly re-plotted together with ac-R(T) of the same sample. Reasonable quantitative concordance between the measured and calculated $R(T)$ is evident from this figure. This finding suggests that, where the crystals' overheating is determined by the heat load, the sample resistance, R=V/I, is a good practical measure of the overheating. 

Moreover, this finding allows an experimental estimate of the characteristic heat loads required to heat the crystal from the arbitrary bath temperature, $T_B$, to a certain target temperature, $T_T$. For example, Fig.1C shows the experimental dependencies obtained for $T_T=100K$ on the same sample in different experimental environments. These curves obey a linear fit (broken lines in Fig.1C), whose slopes depend on the heat transfer efficiency of the set up used.

Fig.1C provides experimental evidence of the applicability of {\it Newton's Law of Cooling} (1701)  to layered HTSC. This result, together with the generic out-of-plane I-V linearity in the normal state of the layered HTSC,  are of principal importance for the self-heating model \cite{heating-1st}, which assumes that heating is the {\it only} cause of non-linearities observed in current-voltage characteristics. Below it will be shown that the majority of results obtained with IT technique could be reproduced using a simplified linear version of the model.

\subsection{The pseudogap, $\Delta_{pg}$}

Let us first address the origin of the pseudo-gap, $\Delta_p$, observed with IT spectroscopy (see, for example,  \cite{pseudogap} and references therein). From a typical example it may be seen that the `tunnelling characteristics' measured in \cite{yurgens2201} over a wide temperature range in the normal state of layered HTSC  could be reproduced {\it quantitatively} using experimental normal state resistance $R(T)$ of the same sample (reproduced in the insert to Fig.2), with the assumption that the mesa heating caused by Joule dissipation is the {\it only} reason for effects observed at high bias (see also \cite{comment2201}). Using  {\it Newton's Law of Cooling} (1701), the temperature of a thin mesa is given by
\begin{equation}
T=T_{B}+ IV/(Ah),
\end{equation}
where $T_B$ is the temperature of the coolant medium (liquid or gas) and $h$ is the heat transfer coefficient. According to Eq.(1), monitoring I-V at a given bath temperature $T_B$ results in a sample temperature rise that entails non-linearity in $I$=$V/R(T)$, which is gap-like if $\partial R/\partial T$$<$0. The $dI/dV$ curves thus constructed resemble those presented in fig.4 of \cite{yurgens2201} and reveal a {\it quantitatively} similar variation of conductance. This similarity allows for an estimate of the heat transfer $(hA)$$\simeq$$16\mu W/K$ and overheating at high bias, $\sim$80K for $T_B$=$200K$; the set of curves accounting for this coefficient is shown in Fig.2. As that Figure clearly shows, the `heating' spectra taken at $T_B$$<$$T^*$ reveal a `pseudo-gap' which disappears entirely as soon as $T_B$$\geq$$T^*$. Interestingly, the IVCs in Fig.2 are nonlinear even above $T^*$. In perfect agreement with the experiment, the {\it heating} spectra have an inverted parabolic shape, which was originally attributed to the presence of van-Hove singularity close to the Fermi level, cf. eg. \cite{pseudogap}. 

Thus, Eq.(1) provides a natural and adequate explanation of the puzzling findings of  \cite{yurgens2201}. Accordingly, the nonlinear I-V characteristics observed by \cite{yurgens2201} in Bi2201 are undoubtedly related to the temperature dependence of the {\it normal} state c-axis resistance of their sample rather than to a (pseudo)gap in the tunnelling density of states (DOS). Moreover, there is little doubt that $R(T)$ modification caused by doping is responsible for the rich zoology of the `pseudogaps' observed in Chalmers \cite{krasnov}.

\begin{figure}
\begin{center}
\includegraphics[angle=-0,width=0.47\textwidth]{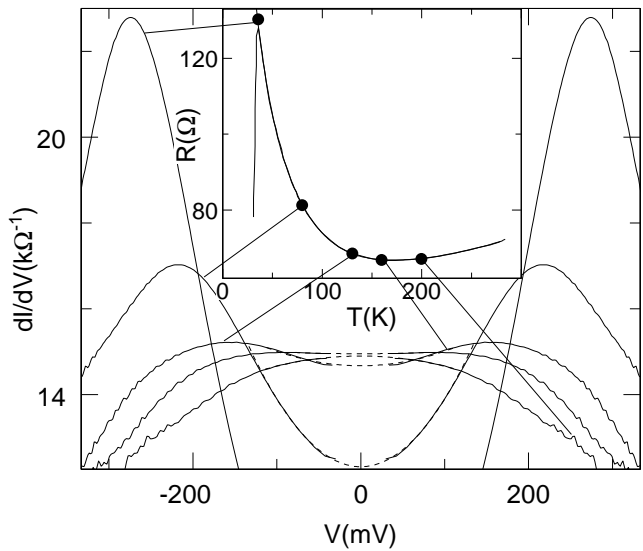}
\vskip -1mm \caption{$dI/dV$ obtained from $R_N(T)$ of \cite{yurgens2201} (shown in the insert) assuming Joule heating origin
of I-V nonlinearities.}  \end{center}
\end{figure}
\subsection{$\Delta_s$ and $\Delta_{pg}$ coexistence}

The validity  of Eq.1 down to 30K, verified in \cite{physicaC}, permits its application to the results originally attributed to the superconducting gap. A lack of information about normal state resistance in the absence of superconductivity, $R_N(T)$ makes this task more problematic. Nevertheless, as was shown elsewhere \cite{physicaC}, the same model provides reasonable agreement with IT spectra measured below $T_c$ using the experimentally verified assumption \cite{heating-1st,comment2201} that the `sub-gap' IT-resistance, $R_{sg}\equiv\partial V/\partial I|_{P\rightarrow0}$, obtained in zero field is a good practical measure of the $R_N$ of a single sample in conditions where its superconductivity is destroyed, for example by sufficiently high magnetic fields. The resulting `heating' $\partial I/\partial V$ curves resemble the experimental ones from \cite{suzuki99} and allow estimation of both the `gap'  and the heat transfer, $(hA)\approx25-33\mu W/K$. The `heating' spectra reveal a gap-like peculiarity whose temperature dependence is remarkably similar to that in \cite{suzuki99}. Hence, Eq.(1) provides a natural explanation of some of the IT-results attributed to the superconducting state \cite{suzuki99}.

%{\small \it In pure mathematics, "well-behaved" objects are those that can be being proved or analyzed by elegant means to have elegant properties. The opposite case is usually labelled pathological}

However, the `heating' IVC reveal {\it one} set of gap-like or S-shaped peculiarities, provided that $R_N(T)$ is a well behaved function as observed \cite{yasuda,my}. Although this agrees perfectly with the majority of IT results, there are a few works which claim that their IT-spectra {\it`evidence the {\it coexistence} of the superconducting gap and the pseudogap'} \cite{pseudogap,krasnov}. At first glance, these observations cast doubts on the model's applicability. More close examination, however, reveals that the gaps' coexistence is observed in 3-probe measurements where contact resistance, $R_c$, inevitably contributes to the net $R(T)$. It is a priori evident that $R_c(T)$ may reveal vastly different behaviour depending on the type of particular contact; see for example \cite{NS}. Besides, $R_c(T)$ is inevitably affected by development of the true superconducting gap of the material. Moreover, for certain types of contacts the contact resistance may become highly non-Ohmic below $T_c$ \cite{pseudogap,NS,textbook1}. For these reasons, the lack of $R_c(I,T)$ curves increases the ambiguity in low temperature behaviour of the net $R(T)$. 

Although the influence of the 3-point configuration on the `heating' IT-spectra might be evaluated for a general case of NS contact of an arbitrary barrier strength using IVC by \cite{NS} together with the normal state $R(T)$ measured in the absence of superconductivity \cite{yasuda,my,my-PRL}, this task falls beyond the scope of this article and will be addressed elsewhere. 

Below I will outline on a qualitative level how the $R_c$ modification caused by the superconducting gap (SG) development might influence the resulting IT spectra. I will show through a simple example that this $R_c$ modification not only affects the net 3-point resistance $R(T)$, but modifies the `heating' spectra {\it radically}. Neglecting $R_c$ nonlinearity for the sake of simplicity, let us assume that the SG contribution to the total contact resistance, $R_c$, in the $T_c$ vicinity might be modelled as $\delta R_c(T)=\delta R_c(0)(1-T/T_c)$. This contribution may cause little effect on the net $R(T)$ provided that $\delta R_c(0)$ is sufficiently small. To visualize this statement, 3-point R(T) curves (broken lines in Fig.3) are constructed from the well-behaved `mother' one (shown by the solid line) for a representative set of $\delta R_c(0)$. For the sake of transparency, this construction uses the same `mother' $R(T)$ as in Fig.2 and assumes $T_c=60K$, as shown by the arrow in Fig.3. At first sight the thus distorted curves seem to differ insignificantly from the undisturbed one even for the highest $\delta R_c(0)=30\Omega$ among those presented in Fig.3. Unlike $R(T)$, the heating spectra are drastically affected by this modification. As is clearly seen in the insert to Fig.3, unlike the `mother' heating spectrum, the spectra modelled for $T_B<T_c$ reveal both a `pseudo-gap', $V_{pg}$, and a `superconducting' gap, $V_s$. 

Thus, this figure reproduces the most significant finding of \cite{pseudogap,krasnov}: it shows that a sharp `superconducting' peak at $V_s$ emerges on top of the hump at $V_{pg}$  associated with the normal state pseudogap in \cite{pseudogap}. The latter feature is evidently a robust signature of the normal state $R(T)$, while the `superconducting' peak is probably caused by the 3-point configuration employed in \cite{yurgens2201,pseudogap,krasnov}. The extrinsic origin of the $V_s$ peak in Fig.3 is additionally illustrated by  the  strong dependence of its amplitude on $\delta R_c(0)$ value. Moreover, more sophisticated analysis has shown the amplitude and position of this peak $V_s$, to be very sensitive to the particular shape of $\delta R_c(T)$ as well as to $R_c$ nonlinearities \cite{practical}. So our model provides a natural explanation of the puzzling findings of  \cite{yurgens2201,pseudogap,krasnov}. However, given the fundamental importance of the conclusions of \cite{pseudogap}, further evidence of the extrinsic origin of the $\Delta_s$ and $\Delta_{pg}$ coexistence is desirable. 

\begin{figure}
\begin{center}
\includegraphics[angle=-0,width=0.47\textwidth]{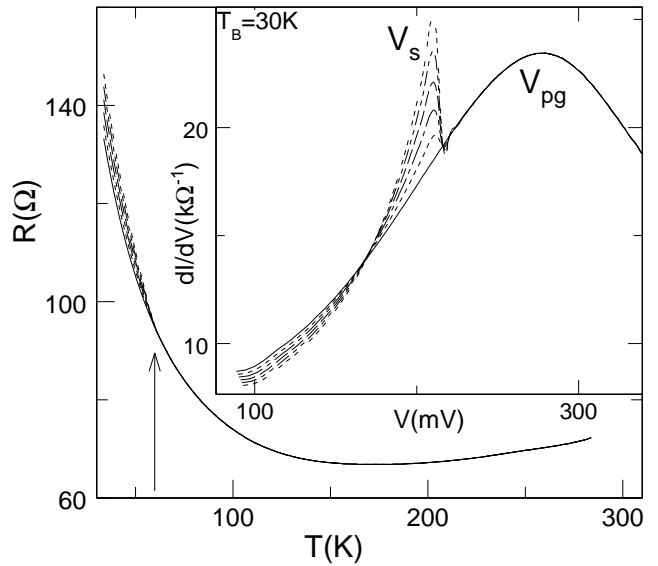}
\vskip -1mm
\caption{Modelling the 3-probe configuration induced coexistence of $\Delta_s$ and $\Delta_{pg}$ within the heating scenario. The solid line represents the `mother' well-behaved $R(T)$ from the inset to Fig.2. Broken lines are constructed from this $R(T)$ assuming that a superconducting gap development below $T_c=60K$ (shown by the arrow) contributes to the contact resistance as $\delta R_c(T)=\delta R_c(0)(1-T/T_c)$ with $\delta R_c(0)=6;12;18;24;30\Omega$ respectively. Insert shows $dI/dV$ obtained f\textbf{}rom the corresponding $R(T)$ curves in the main panel assuming Joule heating origin of I-V nonlinearities. 
}  \end{center}
\end{figure} 

Fortunately, one study, \cite{krasnov}, allows an estimate of the net  resistance of the bridge, thus offering an independent test of our conclusions.   Unlike in the case of direct measurements of normal state resistance in the absence of superconductivity, eg. \cite{heating-1st}, and in qualitative agreement with the a priori expectations mentioned above and illustrated in Fig.3, the resulting 3-point $R(T)$ from \cite{krasnov} reveal a kink-like change  below T$_c$ and a temperature-dependent discrepancy between the normal state R(T) extrapolation towards $T<T_c$  and the low temperature estimates by \cite{krasnov}. In resemblance of the calculated curves in Fig.3, the experimental discrepancy increases rapidly as the temperature lowers further below $T_c$. Remarkably, the relative magnitude of this discrepancy exceeds that in Fig.3 by more than one order, thus additionally supporting our assumptions. Moreover, the spectra constructed with Eq.(1) from the net $R(T)$ by \cite{krasnov} are qualitatively similar to those presented in the insert to Fig.3. In remarkable resemblance of the experimental IVC by the Chalmers group \cite{krasnov}, these spectra reveal both a `pseudo-gap', $V_{pg}$, and a `superconducting' gap, thus again reproducing the most significant finding of \cite{pseudogap,krasnov}: a sharp `superconducting' peak at $V_s$ emerges on top of the hump at $V_{pg}$  associated with the normal state pseudogap in \cite{pseudogap}. 

Unlike the computed curves in Fig.3, the low temperature measurements by \cite{krasnov} reveal noticeable scatter, thus permitting slightly different $R(T)$ approximations. Interestingly, in perfect agreement with \cite{practical}, vastly different amplitudes of the `superconducting' peak {\it and} its positions, $V_s$, were obtained with Eq.(1) from these  $R(T)$ approximations, which are compatible with the error bar of \cite{krasnov}. Unlike the Chalmers group, I believe that the $V_{pg}$ feature is a robust signature of the normal state $R(T)$, while the `superconducting' peak is caused by the 3-point configuration employed in \cite{yurgens2201,pseudogap,krasnov}.

Thus, Fig.3 together with the `heating' spectra calculated from the net 3-point $R(T)$ from \cite{krasnov} provide a natural explanation of the puzzling findings of \cite{krasnov}. Interestingly, the heat transfer efficiency $(hA)$$\simeq$$8.4\mu W/K$ and overheating at high bias, $>$200K for $T_B$=$60K$ estimated for the case of \cite{krasnov} correlate reasonably with the estimates obtained in the previous section.  Moreover, complementary to the `heating' spectra, our model allows reconstruction of $R(T)$ dependence from intrinsic tunnelling I-V obtained at fixed $T_B$. Remarkably, the same $hA$ allows  perfect superposition of the $R(T)$ thereby  constructed  with Eq.(1) from I-V by \cite{krasnov} with experimental $R(T)$ of the same sample. Thus, the results of this section demonstrate the self-consistency of the model. 

Hence, it is most probably fallacious to identify  the symmetric peaks in IT-spectra with the superconducting energy gap $\Delta_s$ or  normal state pseudo-gap $\Delta_p$ in the electronic density of states (DOS). Indeed, experimental IT spectra undoubtedly originate from Joule heating and are related to the peculiar temperature dependence of the net resistance of the bridge. The three-point configuration developed in Chalmers is most likely responsible for the resulting peculiar if not pathological $R(T)$.  The `heating' spectra might of course also be affected by the contacts' I-V deviations from linearity. However, this contribution cannot be verified reliably because of the lack of contact characterisation in \cite{pseudogap,krasnov}.

\subsection{Magnetic field influence on $\Delta_s$ and $\Delta_{pg}$}

Let us turn now to the puzzling effect of the magnetic field on the IT features ascribed to the superconducting gap, $V_s$, and the c-axis pseudogap, $V_{pg}$. Recently, Krasnov \cite{pseudogap} observed that $V_s$ and $V_{pg}$ have  different magnetic field dependencies. Ignoring self-heating, Krasnov considered these observations to represent a crucial test for the superconducting origin of the two gaps in DOS of HTSC. However, the thermal load estimate even for HgBr$_2$-Bi2212 crystal at Helium temperature exceeds KILOWATTS/cm$^2$, so that this assumption is probably incorrect.
  
Moreover, I will show below that the same model allows a natural explanation of this IT finding, namely the contrasting effect of the magnetic field, $B$, on $V_s$ and $V_{pg}$, where the first characteristic was found to be much more sensitive to the field \cite{pseudogap,krasnov-B}. In my opinion, the normal state magnetoresistance (MR) of the substance investigated and its temperature dependence \cite{my} are responsible for that finding. Indeed, as shown above, to reach $V_s$ and $V_{pg}$ features, one has to warm the sample almost up to $T_c$ and $T^*$, the minimum in R(T), respectively. The negative normal state MR in Bi2212 is experimentally found to be small at $T\sim T^*$ \cite{my,my-PRL}, thus ensuring negligible $V_{pg}(B)$ dependence, in perfect agreement with the experiment \cite{pseudogap,krasnov-B}. However, the magnitude of the normal state MR grows rapidly on temperature lowering \cite{my,my-PRL}, so that the noticeable field influence on $V_s$ is also explained naturally within the heating scenario\cite{heating-1st,comment2201}. 

Fortunately, the data of \cite{krasnov-B} allow a cross-check of the model predictions, since 3-point $R(T)$ curves are presented for a set of fields. Of course, the unknown $R_c(B,T,I)$ in \cite{krasnov-B} prevents any reliable MR estimate for the HTSC investigated. However, these curves allow an estimate of the net 3-point MR which is responsible for the magnetic field influence on the IVC. This MR is huge, 30\%(@14Tesla) at 4K and it drops below the resolution of \cite{krasnov-B} as the temperature increases towards $T_c$, as seen in the insert to Fig.4. Evidently, these $R(T)$ curves allow estimation of the effect of magnetic field on the `heating' IT spectra. Fig.4 shows dI/dV spectra obtained with Eq.(1) from R(T) measured at two fields. Qualitative similarity with the results of \cite{pseudogap,krasnov-B} is evident from Fig.4. Indeed, the heating spectra in Fig.4 evidence gap suppression by magnetic field and temperature in reasonable agreement with the experiment \cite{pseudogap,krasnov-B}. I believe this similarity favours the self-heating origin of the results obtained in Chalmers especially given the unknown field, current and temperature dependence of their contact resistance, $R_c$. 

\begin{figure}
\begin{center}
\includegraphics[angle=-0,width=0.47\textwidth]{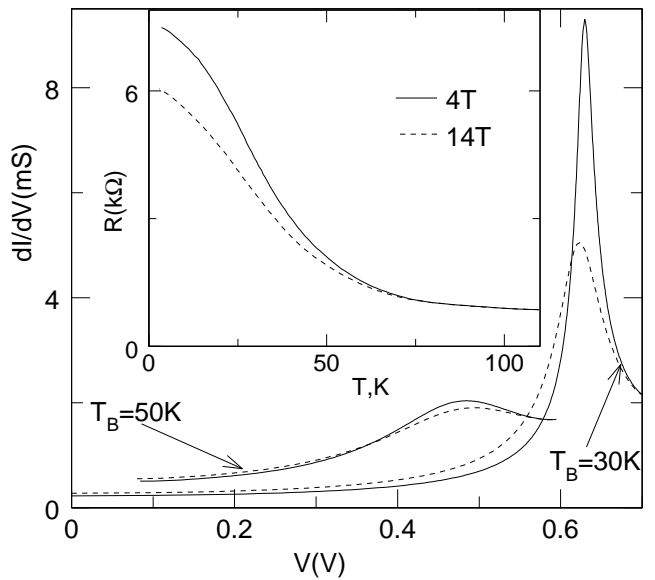}
\vskip -1mm
\caption{ $dI/dV$ obtained from $R(B,T)$ of \cite{krasnov-B} (shown in the insert) assuming Joule heating origin of I-V nonlinearities. %\cite{krasnov}. 
}  \end{center}
\end{figure} 

Interestingly, the heat transfer efficiency $(hA)$$\simeq$$8.5\mu W/K$ estimated for this experiment is strikingly  similar to other cases considered here, which provides extra support for the model. So the most significant findings of high-bias ITs could be broadly understood within the heating scenario.  

In conclusion, temperature dependent normal state magnetoresistance is the most likely cause of the contrasting response to magnetic field of the two gaps observed in intrinsic tunneling experiments \cite{krasnov-B,pseudogap}. This provides a natural explanation of the $V_s$ closure well below $T_c$ of the material as well as its survival at magnetic fields which significantly exceed $H_{c2}$, the upper critical field of the material \cite{comment-1st,my,my-PRL,alezav,my-MR}. Although the low-bias results reported by this group might be less affected by self-heating \cite{physicaC,heat-free}, these are of little relevance because of the 3-point configuration and unknown field-temperature-current dependence of the contact resistance employed in \cite{yurgens2201,krasnov-B,pseudogap}.  

\subsection{Low-bias IT data}
Undoubtedly, there is a certain level of dissipation at which intrinsic phenomena may dominate heating. Extra experiments are required in order to address this issue unequivocally. However, certain preliminary conclusions may be drawn from analysis of the so-called `low-bias' IT-curves. To simplify the task, let us consider IT results obtained on the Dayem-bridge-like HTSC-structures. These structures benefit from spatial separation between the junction and the contacts, since the Joule heating of the contacts might influence the mean temperature of the junction less than in the case of the stacked geometry of mesas considered in the previous sections. Thus, these objects are highly suitable for the purpose of this section. 

Similar to other 4-point IT-studies, the current-voltage characteristics of these bridges reveal {\it one} gap-like peculiarity, $V_s$  and exceptionally well defined and nearly temperature-independent Ohmic resistance at very high bias, which was originally attributed to the normal-state tunnel resistance, $R_n$. However, as was first noted in \cite{comment-1st}, the very high levels of dissipation required to achieve these characteristic features at helium temperatures, namely, 7.3 and 210kW/cm$^2$ respectively, evidence the heating origin of these features. Fortunately, complementary to the high-bias IVC, the authors of \cite{latyshev} also measured a low-bias I-V (see Fig.5A) which may be less affected by heating. It will be shown below that our model allows discrimination between regimes of strong self-heating and those where intrinsic phenomena may dominate heating. It will be seen that our model allows  estimation of the highest (critical) heat load that this particular sample can bear at $T_B$ without noticeable overheating. Thus, it allows discrimination between regimes of strong self-heating and those where intrinsic phenomena may dominate heating.

\begin{figure}
\begin{center}
\includegraphics[angle=-0,width=0.47\textwidth]{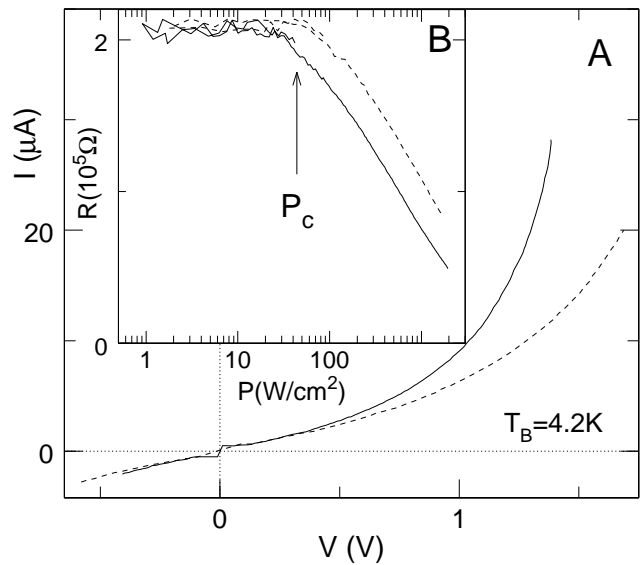}
\vskip -1mm
\caption{{\bf A:}  low-bias I-V of the bridge \#2 from Fig.1b and 3 of \cite{latyshev} are shown by the solid and broken lines respectively; {\bf B:} heat load dependence of bridge \#2 resistance, R=V/I, obtained from Fig.5A.}  \end{center}
\end{figure} 

Evidently, monitoring I-V at a given base temperature, $T_B$, results in a sample temperature rise as soon as $P$ exceeds the critical heat load $P_c$. Indeed, as is seen from Fig.5B, there is a well defined threshold level, $P_c$, of the heat load below which $R(P)$ is flat, while it drops rapidly at $P\geq P_c$. The $R(P)$ curves in the latter case are probably caused by Joule self-heating. Indeed, similarly to other IT examples considered here, these curves resemble R(T) by \cite{latyshev} and reveal a {\it quantitatively} similar variation of resistance, which allows an estimate of the  heat transfer coefficient $h$$\sim$30-60Wcm$^{-2}$K$^{-1}$ (the scatter reflects the discrepancy between Fig.1b and 3 in \cite{latyshev} which is also clearly seen in Fig.5A here). The set of curves accounting for this coefficient correlate reasonably with the measured $R(T)$ curves, thus confirming the heating origin of the falling part of $R(P)$ dependence in Fig.5B. Thus, Eq.(1) provides a natural explanation even of the `low-bias' findings of \cite{latyshev} and suggests that the `heating-free' I-V is linear, in agreement with the conclusions drawn from the analysis of IVC measured over an extended range \cite{physicaC}.

To conclude, the character of the Joule heating of a sample with typical c-axis R(T) dependence is responsible not only for $V_s$ and $R_n$ but also for the majority of the {\it low-bias data} by \cite{latyshev}. Moreover, our model allow evaluation of the critical heat load below which intrinsic phenomena may dominate heating.  The linearity of the thereby determined intrinsic (heating-free) response of Bi2212 from \cite{latyshev} suggested by Fig.5B has little in common with the Josephson behaviour taken for granted by IT advocates.

\subsection{Some more smoking guns for the model}

The self-heating model shows that the vast variety of results obtained with {\it intrinsic tunnelling spectroscopy}  most certainly originate from Joule heating of the sample and are related to the temperature dependence of {\it normal} state resistance. The remarkably similar heat transfer coefficients estimated here for various Bi2212 and Bi2201 microbridges explored by different groups worldwide provide additional support for the model. Moreover, there are several more `smoking guns' which demonstrate that the model is correct. 

In particular, Eq.(1) suggests that, providing $h$ is constant, mesas with different areas would require the same heat loads to achieve the same I-V features. Indeed, strikingly similar loads are required to achieve the characteristic features of the I-V of Bi2212 mesas of a vastly different area, 25-80$ \mu m^2$,  at the same base temperature, 4.3K namely, $3.3kW/cm^2$ and $18.5kW/cm^2$ for those those ascribed in  \cite{krasnov} to the superconducting gap, $V_s$, and to the normal state Ohmic response, $R_n$, respectively. 

Magnetic field offers one more test of the model, since it modifies R$_N$ significantly \cite{my,my-PRL}, while the heat-transfer channel remains virtually insensitive to the field. So, provided the intrinsic heating model is valid, one has to expect the heating-induced features of I-V, and the critical heat load in particular, to be field independent. The `isothermal' I-Vs taken on the Bi2212 Dayem bridge at different fields (inset to Fig.4 in \cite{morozov} is reproduced in Fig.6A here) offer an independent test of the model. These data reveal a quasi-linear negative magnetoresistance, whose value, $\delta R_{sq}(30T)/R_{sq}(0)\simeq -42\%$, correlates reasonably with direct measurements of the normal state MR in Bi2212 \cite{my,my-PRL,my-MR}. Fig.6C shows the same data re-plotted as a sample resistance, R=V/I, normalised by its value at V$\rightarrow$0  versus the heat load, P=IV/A. With remarkable similarity to Fig.5B, these curves reveal a well defined threshold level $P_c$, which separates the intrinsic and extrinsic contributions. Moreover, as is clear from Fig.6C,  different I-V merge onto the same curve, thus supporting the Joule heating origin of the deviations from I-V linearity. 

\begin{figure}
\begin{center}
\includegraphics[angle=-0,width=0.47\textwidth]{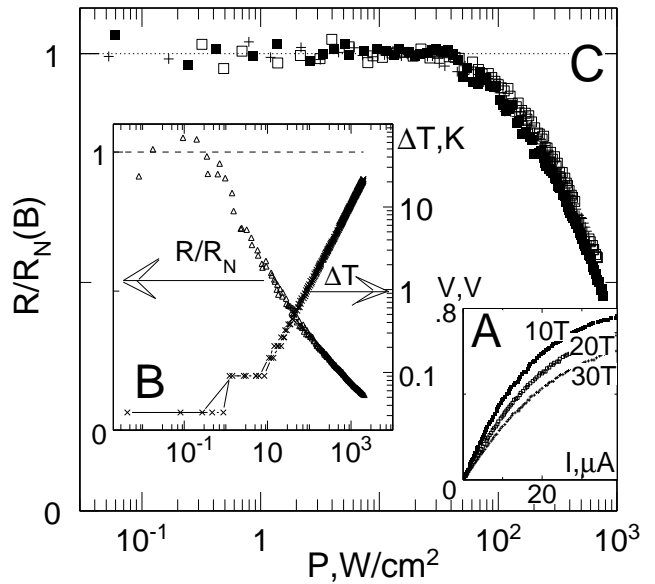}
\vskip -1mm
\caption{{\bf A} The `isothermal' IVC's of the Bi2212 Dayem bridge taken at at different fields from \cite{morozov}; {\bf C} The heat load dependence of the Dayem bridge' resistance, R=V/I, normalised by $R_N(B)$  for 10, 20, and 30Tesla ($\blacksquare,\square,+$ respectively). Insert {\bf B} shows mutual correlation between the mean sample heating and deviations from I-V linearity from\cite{heat-free}.}  \end{center}
\end{figure} 

Moreover, one may note the qualitative similarity with the results of a similar analysis obtained for Bi2212 mesas \cite{heat-free}, reproduced in Fig.6B. Here, the resistance of the sample is plotted together with the topmost electrode overheating above the base temperature, $T_B$. Fig.6B suggests the Ohmic response (the broken lines in Fig.6B) to occur in the absence of overheating. At higher heat loads, we see a progressive rise in overheating, which entails a corresponding reduction of $R_N$.  
The dotted line in Fig.6B represents the intrinsic `heating-free' I(V), which has little in common with Josephson behaviour. The noticeable difference between the critical heat loads estimated from the point of I-V departure from linearity for Bi2212 mesas and 3D-microbridges is probably due to differences in the heat transfer efficiency in these experiments. Besides, $P_c$ for the first case is most likely underestimated because of the unknown contact resistance.

\subsection{How does the heating-free I-V look?}

The results presented in Fig.5B,6B and especially the I-V scaling in Fig.6C provide a strong indication that the departures of experimental I-V from linearity originate in intrinsic heating. These results suggest the intrinsic (`heating-free') out-of-plane response to be Ohmic. Such a conclusion drastically contradicts the naive picture of  Josephson interplane tunnelling \cite{latyshev} but is supported by direct experiments \cite{my_i-v,practical} performed sufficiently close to $T_c$ or the irreversibility field, $H_{irr}$. 

In particular, comparison of the zero-field I-V measured  at the same base temperature on a single sample situated in a vacuum chamber and immersed in liquid nitrogen reveals highly nonlinear Josephson-like branches in the first environment and practically linear ones in the latter case. I believe that the linear branches in this experiment are measured at an insignificant level of Joule heating of the structures because (i) the heat transfer coefficient into liquid radically exceeds that into a vacuum and (ii) the characteristic heat load measured for these branches falls well below the threshold $P_c$ estimate. Remarkably similar results were obtained from the systematic mixed-state I-V studies performed in the vicinity of $H_{irr}(T)$ at vastly different orientations of the magnetic field with respect to the (ab)-plane, as is illustrated in Fig.7.

\begin{figure}
\begin{center}
\includegraphics[angle=-0,width=0.47\textwidth]{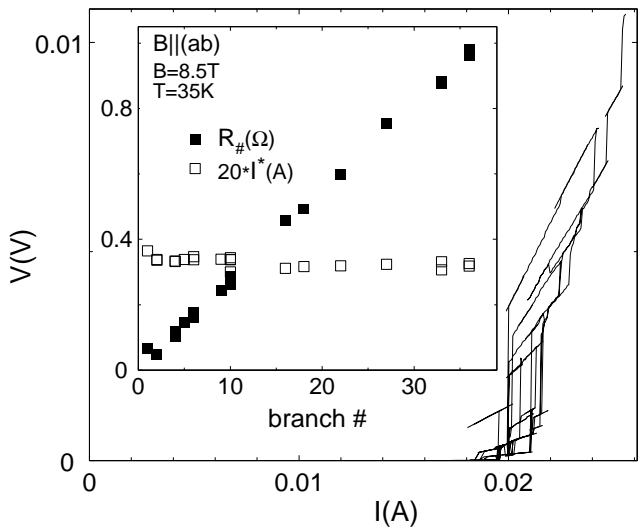}
\vskip -0.5mm
\caption{Typical I-V brush measured under conditions where the critical parameters of Bi2212 are suppressed by magnetic field $\parallel$(ab). Insets show corresponding differential resistances, $R_\#$, and  excess current, $I^*$, of the branch number \#. }
\end{center}
\end{figure}

This similarity makes it possible to outline the distinctive features of the brush-like I-V measured at negligible self-heating: (i) a voltage jump takes place from zero voltage at some characteristic current, $I_J$; (ii) all resistive branches have approximately the same excess current, $I^*$, (given by the intersection of their slopes with the current axis); (iii) the differential resistance $R_\#$ of a resistive branch is proportional to its number, \#, and represents a fraction of the normal state resistance of the same junction measured under conditions of complete suppression of its superconductivity. The remarkable similarity of these features with the characteristic manifestations of phase-slip centers of different dimensionality \cite{phase-slip,Bezryadin,ustinov} in conventional superconductors suggests the phase-slip mechanism might be responsible for the `intrinsic Josephson effect'.

 Evidently, the typical geometrical shape of a Bi2212 sample is a thin flake rather than the quasi one-dimensional samples employed in classical PSC studies. However, those quasi-1D samples were made of almost isotropic material, in drastic contrast to Bi2212, which reveals extremely high and temperature-dependent electric anisotropy, $\rho_c/\rho_{ab}>10^5$, \cite{my-MR}. On taking account of this anisotropy, one immediately concludes that the typical length-to-width ratio for an `isotropic' Bi2212 sample falls into the 10-1000 range (i.e. is the same if not bigger than the typical ratios quoted in classical PSC studies \cite{phase-slip,Bezryadin}), so that our Bi2212 samples could be treated as quasi-1D. Nevertheless, Bi2212 seems to fall beyond the scope considered in \cite{phase-slip} because the width of the sample exceeds both the coherence length and the magnetic field penetration depth. However, the most recent experiment \cite{ustinov} shed some light on the problem. The authors of \cite{ustinov} reported behavior very similar to PSCs (and to our findings) to occur in wide superconducting films (of the width $d$), in which both the coherence length  and the magnetic field penetration depth  are significantly smaller than $d$. In contrast with the 1D case, it was found that oscillations of the order parameter may not necessarily be uniform along the two-dimensional analogue of a PSC. In particular, it was shown that these oscillations may occur in the form of propagation of waves carrying the order parameter singularities across the film. It seems plausible that likewise \cite{phase-slip,Bezryadin,ustinov} nonequilibrium superconductivity of the phase-slip planes (PSP) might be responsible for the heating-free brush-like I-V reported here.

In fact, the one-dimensional phase-slip center could naturally be considered as a subclass of weak links, characterised by direct conductivity (unlike that of well-known tunnel junctions). Such weak links have been extensively studied, particularly for the practical realisation of Josephson electronics, cf. eg. review \cite{licharev}. Contrary to tunnel junctions, weak links are characterised by a high current density, which generates a highly nonequilibrium state in the structure. A quantitative description of this state is rather complicated because it must address mutually correlated nonequilibria in three systems: the condensate, the quasiparticles and the phonons of the structure under consideration, together with its interaction with the environment. Despite the absence of any selfconsistent theoretical description, the present-day level of phenomenological understanding \cite{textbook1,phase-slip,Bezryadin,ustinov} appears sufficient to obtain a general idea about the underlying processes. Let us consider a model sample containing one (or many) PSC and account for the Joule heat generated in such a center. This heat can only escape through the surface of the sample by thermal conductivity along the {\it superconducting} material and/or directly into the substrate (or rather to the exchange gas). Heating is small when there is sufficiently good thermal contact and moderate Joule dissipation. In this case, the differential resistance of PSC is known to be determined by the normal state resistance, $R_N$, \cite{phase-slip} which would be obtained on a single sample in conditions where its superconductivity is destroyed somehow, for example by magnetic fields \cite{physicaC}. In most cases, however, the thermal contact is far from ideal and this entails more or less noticeable heating of the area near the PSC. If the dissipation in the phase-slip centre raises its temperature {\it above} the transition temperature, a normal `hot spot' \cite{gurevich} is created and an SNS junction is therefore formed. The Josephson effects are eventually quenched when the normal region grows. 

Surprisingly, this oversimplified scenario seems to be compatible with the experiment. Indeed, the differential resistance of the `heating-free' branch correlates reasonably with $R_N$ \cite{heating-1st}. The increasingly strong deviations from linearity with the heat load correspond naturally to a case in which heating of the PSP becomes noticeable on the scale of $R_N(T)$. As soon as the bias is sufficiently high, either the PSP or the whole sample becomes normal; however, its I-V remain tunnel-like and reveal a clear `gap' until the mean temperature of the stack exceeds that of the minimum in $R_N(T)$, in agreement with \cite{physicaC}. Finally, at exceptionally high levels of dissipation and heating, the `pseudo-gap' humps disappear and we obtain a slightly convex I-V, which is frequently identified with $R_N$. The mean temperature of the overheated stack, naturally, gives a very coarse estimate of the highly nonequilibrium temperature {\it distribution} across the stack and should therefore be treated with great caution \cite{physicaC}.

\section{conclusions}The intrinsic heating model proposed in \cite{heating-1st} is constructed on the basis of Newton's Law of Cooling and the linearity of I-V experimentally obtained in the normal state of the crystal. This model shows that the vast variety of results obtained with {\it intrinsic tunnelling spectroscopy} originate from the Joule heating of the sample and are related to the temperature dependence of {\it normal} state resistance. In particular, the IT feature attributed to the normal state pseudogap in the electronic density of states is an indirect though robust signature of the true normal state $R(T)$, while the `superconducting' one is most likely caused by the 3-point configuration developed in \cite{yurgens2201,pseudogap,krasnov}. Normal state magnetoresistance is the most probable cause of the contrasting response to magnetic field of these two gap-like I-V peculiarities. Moreover, this scenario naturally explains why the  `superconducting' peak in \cite{yurgens2201,pseudogap,krasnov} collapses  well below the superconducting critical temperature, $T_c$, of material as well as why it survives a magnetic field which significantly exceeds the upper critical field of this substance; see \cite{comment-1st,my,my-PRL,my-MR} for more detail.  Thus, the conclusions gathered by temperature, magnetic field, intercalation and doping dependent IT are most likely irrelevant. Unfortunately, even the low-bias results reported by this group are of little help because of the 3-point configuration and unknown field-temperature-current dependence of the contact resistance. Indeed, analytical and experimental results suggest that, provided self-heating is negligible, the individual branches of the brush-like I-V are linear. This finding does not support the conventional picture of `intrinsic' Josephson tunnelling. Moreover, it seems plausible that, as in \cite{phase-slip,Bezryadin,ustinov}, stationary and nonstationary nonequilibrium superconductivity, phase-slip planes for example, might be responsible for the heating-free brush-like out-of-plane I-V in HTSC.

\section{acknowledgements} I gratefully acknowledge the financial support of the Leverhulme Trust (F/00261/H), stimulating discussions with A.S.Alexandrov and A.V. Ustinov, valuable suggestions by the anonymous referee of Ref.\cite{comment2201}, and Bi2212 mesa manufacturing by A.Yurgens. I also thank G.Grabtree for his interesting contribution.

\end{document}